\documentclass{myjfm}

\usepackage{graphicx,color}
\usepackage{epsfig}
\usepackage{natbib}
\usepackage{bm}
\usepackage{amsmath}
\usepackage{color}

\ifCUPmtlplainloaded \else
  \checkfont{eurm10}
  \iffontfound
    \IfFileExists{upmath.sty}
      {\typeout{^^JFound AMS Euler Roman fonts on the system,
                   using the 'upmath' package.^^J}%
       \usepackage{upmath}}
      {\typeout{^^JFound AMS Euler Roman fonts on the system, but you
                   dont seem to have the}%
       \typeout{'upmath' package installed. JFM.cls can take advantage
                 of these fonts,^^Jif you use 'upmath' package.^^J}%
      }
  \else
  \fi
\fi

\ifCUPmtlplainloaded \else
  \checkfont{msam10}
  \iffontfound
    \IfFileExists{amssymb.sty}
      {\typeout{^^JFound AMS Symbol fonts on the system, using the
                'amssymb' package.^^J}%
       \usepackage{amssymb}%
         \let\leq=\leqslant
         
      }{}
  \fi
\fi

\ifCUPmtlplainloaded \else
  \IfFileExists{amsbsy.sty}
    {\typeout{^^JFound the 'amsbsy' package on the system, using it.^^J}%
     \usepackage{amsbsy}}
    {}
\fi

\newsavebox{\astrutbox}
\sbox{\astrutbox}{\rule[-5pt]{0pt}{20pt}}

\title[Inertial particles in a VK flow ]{Dynamics of inertial particles in a turbulent von K\'arm\'an flow}

\author[R. VOLK et al.]{R. VOLK, E. CALZAVARINI, E. L\'EV\^EQUE, J.-F. PINTON}

\affiliation{International Collaboration for Turbulence Research\\
Laboratoire de Physique de \'Ecole Normale Sup\'erieure de Lyon, UMR5672 \\
CNRS et Universit\'e de Lyon, 46 All\'ee d'Italie, 69007 Lyon, France.}  
\date{\today} 

\begin{document}
\maketitle

\begin{abstract}
We study the dynamics of neutrally buoyant particles with diameters varying in the range $[1, 45]$ in Kolmogorov scale units  ($\eta$) and Reynolds numbers based on Taylor scale ($Re_{\lambda}$) between 580 and 1050. One component of the particles' velocity is measured using extended Laser Doppler Velocimetry at the center of a Von-Karman (VK) flow, acceleration is derived by differentiation.
We find that although the particle acceleration variance decreases with increasing their diameter with scaling close to $(D/\eta)^{-2/3}$, in agreement with previous observations, the characteristic time of acceleration autocorrelation increases much strongly than previously reported, and linearly with $D/\eta$. A new analysis of the probability density functions of the acceleration shows smaller wings for larger particles; the flatness indeed decreases as also expected from the behavior of Eulerian pressure increments in the VK flow. We contrast our measurements with former observations in wind-tunnel turbulent flows and numerical simulations, and discuss if the  observed differences arise from inherent properties of the VK flow.
\vskip 0.2cm 
\end{abstract}

%\begin{keywords}
%Turbulence, Inertial Particles, Intermittency
%\end{keywords}

%%%%%%%%%%%%%%%%%%%%%%%%%%%%%%%%%%%%%%%%%%%

\section{Introduction}
%%%%%%%%%%%%%%%%%%%
Research in dynamics and transport phenomena in turbulence have recently benefitted from experimental tracking of flow tracers, see for instance (\cite{Ott:2000,laporta:2001,mordant:2001,arneodo:2008,toschi:2009}). Ideally these tracers should have a size much smaller than the Kolmogorov length $(\eta)$ where the velocity gradients are smooth and hence their motion follows fluid streamlines, but experimental constraints have often lead to the use of larger particles -- with some bias as discusses {\it e.g.} in (\cite{mei:1996,Voth:2009}).  On the other hand, the question of the dynamics of  objects with a finite size freely advected by turbulent motions is a question on its own. Theories developed in the  small particles limit at vanishing particle Reynolds numbers $Re_p$ yield the celebrated Maxey-Riley-Gatignol equation, with little counter part for high-$Re_p$ situations (see, however, ~\cite{auton:1988, Lovalenti:JFM:1993, Loth:EFM:2009}). A recent systematic analysis has been made in a wind tunnel ($Re_{\lambda} = 160$) using helium inflated soap bubbles (\cite{qureshi:2007}).  Other studies were performed with isodense polystyrene particles in water in a turbulent Von-Karman flow ($Re_{\lambda} \in [400, 815]$)(\cite{Voth:2002,Voth:2009}). 
They  have obtained several noteworthy results: 
(i) the variance of acceleration decreases as $D^{-2/3}$, after the manner with which pressure increments vary with size according to the Kolmogorov scaling prediction. The influence of varying Reynolds numbers ($Re_{\lambda} \in [400, 815]$) has also been studied in \cite{Voth:2009} showing that the variance of acceleration actually scales according to $\epsilon^{3/2} \nu^{-1/2} (D/\eta)^{-2/3}$. 
(ii) The probability distributions functions (PDFs) of acceleration components do not depend on particle sizes in the explored range $D/\eta \in [12, 25]$ \cite{qureshi:2007} and  $D/\eta \in [0.4, 27]$ (\cite{Voth:2009}). \cite{Voth:2009} note that PDFs for large particles sizes may be slightly below the fluid particle PDF however due to systematic uncertainties they can not draw any firm conclusion. Our numerical study based on the Fax\'en Model  (\cite{Calzavarini:JFM:2009}) suggests instead that the PDF shape should become narrower at increasing the particle size for $Re_{\lambda} = 75, 180$. This finding is questioned by a more recent numerical study (~\cite{Homann:arxiv:2009}) based on a direct simulation approach via penalty methods, which finds a collapse of the PDFs for $D/\eta \in [2,14]$ at $Re_{\lambda} = 32$.{We go beyond the above mentioned observations using new experimental data obtained by extended Laser Doppler Velocimetry in a VK flow (\cite{volk:2008a}). We find that: (i) There may be corrections to the  $(D/\eta)^{-2/3}$ scaling of acceleration variance which comes from intermittency. (ii) The acceleration PDFs normalized by their variance are not independent of the particle size. By an indirect statistical analysis we observe that the wings of the distributions become less extended at increasing $D/\eta$. (iii) The response time of the particle, as computed from the acceleration autocorrelation function, increases much strongly than previously reported in \cite{Calzavarini:JFM:2009}, and linearly with $D/\eta$. (iv) Still there are now consolidate differences between results in von K\'arm\'an (VK) and wind-tunnel (WT) turbulence, such as different functional forms for the acceleration PDFs, or the trend in the acceleration correlation time with particle size. We argue that such discrepancies may originate from the large scale structure and anisotropy characterizing the VK flow.}
\section{Experimental setup} 
%%%%%%%%%%%%%%%%
The flow is of the von K\'arm\'an type, as described in~\cite{volk:2008b}.  Water fills a cylindrical container of internal diameter 15~cm, length~20 cm. It is driven by two disks of diameter 10 cm, fitted with blades in order to increase steering. The rotation rate is fixed at values up to 10~Hz. For the measurements reported here, the Taylor based Reynolds number reaches up to 1050 at a maximum dissipation rate $\epsilon$ equal to 22 W/kg. The flow temperature is regulated to be constant and equal to $15^{\circ}$C at all rotation rates. The tracked particles have a density of 1.06, with diameters $D=30, 150, 250, 430, 750 \; \mu{\rm m}$. When further changing the flow stirring, this will corresponds to $D/\eta  \in [1, 45]$. 

\begin{table}%[h!]
\begin{center}
\begin{tabular}{|c c c c c c c c|}
\hline  $\Omega$ & $u_{\rm rms}$ & $a_{\rm rms}$ & $\tau_\eta$ & $\eta$ & $\epsilon$ &  $R_\lambda$ & $a_0$\\
\hline  ${\rm Hz}$ & ${\rm m.s}^{-1}$ & ${\rm m.s}^{-2}$ & ms & $\mu$m & ${\rm W.kg}^{-1}$ &  - & - \\
\hline  4.1 & 0.57 & 144 & 0.53 & 24.8 &  4  & 580 & $2.8$ \\
\hline  6.4 & 0.85 & 375 & 0.33 & 19.6 &  10.2  & 815 & $4.6$ \\
\hline  7.2 & 0.99 & 496 & 0.28 & 18.2 & 13.9  & 950 & $5.1$ \\
\hline  8.5 & 1.17 & 706 & 0.29 & 16.2 & 21.8  & 1050  & $5.2$ \\
\hline
\end{tabular}
\end{center}
\caption{Parameters of the flow. $\Omega$: rotation rate of the disks, $\epsilon$  dissipation rate, from the power consumption of the motors (accuracy of about $20\%$). The Taylor-based Reynolds number is computed as $R_\lambda=\sqrt{{15 u_{rms}^4}/{\epsilon\nu}}$, and $a_0$ is derived from the Heisenberg-Yaglom relation {$a_0 \equiv a_{rms}^2\  \nu^{1/2} \epsilon^{-3/2}$}.}
\end{table}

Particles are tracked using the extended Laser Doppler Velocimetry introduced in~\cite{volk:2008b,volk:2008a}. We use wide Laser beams to illuminate particles on a significant fraction of their path. When a particle crosses the fringes, the scattered light is modulated at a frequency directly proportional to the component of the velocity perpendicular to the fringes. As the beams are not collimated, the inter-fringe remains constant across the measurement volume whose size is about $5 \times 5 \times 10 \textrm{mm}^3$. In practice, we use a 2W continuous Argon laser of wavelength 514nm to impose a $41 microns$ inter-fringe, and image measurement volume on a low noise Hamamatsu photomultiplier in the case of the smallest (fluorescent) particles, while for larger particles, the detection is made using a PDA-36A photodiode from Thorlab. The output is recorded using a National Instrument PXI-NI5621 16bit-digitizer at rate $1$~MHz. The velocity is computed from the light scattering signal using the demodulation algorithm described in~\cite{Mordant:PhyD:2002}, with a time resolution of about 30~$\mu$s. The acceleration is then computed by differentiation of the velocity output. Because of measurement noise the signal has to be filtered using a gaussian smoothing kernel with window width $w$ as proposed in~\cite{Mordant:PhyD:2004}. Moments of the statistics of fluctuations of acceleration are computed for varying values of $w$ and interpolated to zero filter width.

\section{Results}
%%%%%%%%%%
\subsection{Particle velocities}
%%%%%%%%%%%%%%%%%
One expect that Eulerian and Lagrangian velocity statistics coincide under ergodicity approximation, so that the tracer particles velocities are expected to have Gaussian statistics.  
\begin{figure}
\begin{center}
  \includegraphics[width=6.5cm]{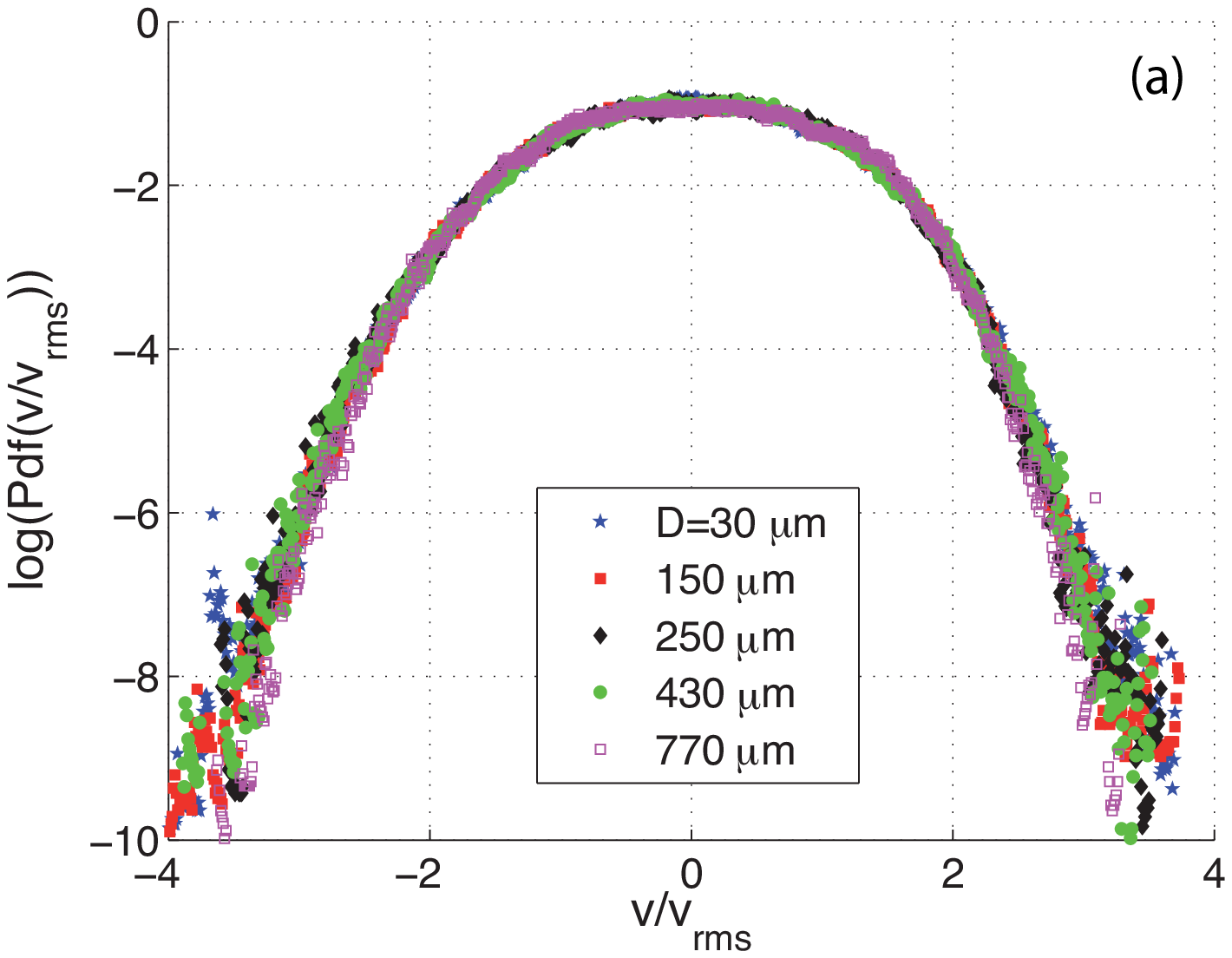} 
   \includegraphics[width=6.5cm]{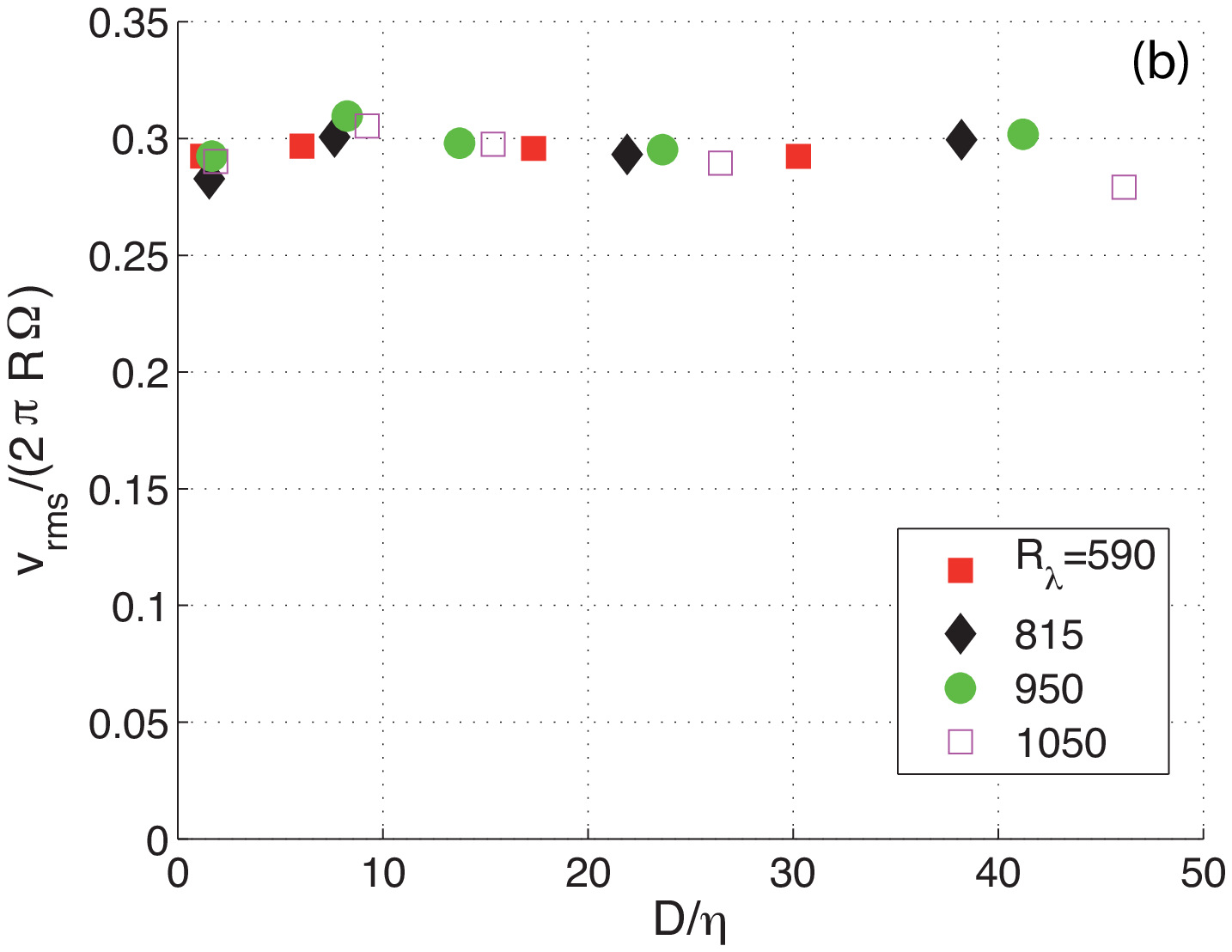}
\end{center}
\caption{(a) Velocity statistics; (b) Evolution of variance with particle size.}
\label{fig1}
\end{figure}
Our observation is that the velocity distribution is markedly sub-Gaussian as seen in figure~\ref{fig1}(a) -- flatness values are ($2.56, 2.58, 2.62, 2.46$) for the 4 Reynolds number values explored in this work. Sub-Gaussian statistics for the velocity have been observed in many experimental set-ups, however usually less pronounced than in our case. 
{Flatness values for velocities in WT flow are closer to 3 (Bourgoin private communication).}
Here the von K\'arm\'an flow has a large scale inhomogeneity and anistropy ({\it c.f.}~\cite{Marie:POF:2004,Ouellette:NJP:2006,Volk:POF:2006})  which may enhance the sub-gaussianity. In such a confined geometry indeed, the VK flow is known to have several possible configurations of its large scale velocity profile (\cite{Monchaux:PRL:2006,Torre:PRL:2007}); each configuration may lead to Gaussian velocity fluctuations about a locally different mean value with an overall effect leading to a sub-Gaussian histogram. However, we have not observed any change in the velocity statistics when the Reynolds number is increased or when the the size of the particle is changed by over an order of magnitude in $D/\eta$. In fact, for this fully turbulent regime, the velocity variance is equal to 30\% of the impeller tip speed -- \ref{fig1}(a). This is a characteristic of the von K\'arm\'an driving impellers, and not a characteristic of the inertial particle size, {\it c.f.}~\cite{Ravelet:JFM:2008}. We note that this observation is in agreement with a prediction following Fax\'en argument at the leading order, $v^2 /u^2_{\rm fluid} = 1 - (1/100)(D^2/\lambda^2)$ (\cite{Homann:arxiv:2009}) -- where $\lambda$ is Taylor's microscale, {giving a correction smaller than $1\%$ at the Taylor-Reynolds numbers considered here}.% {The Fax\'en signature in the velocity statistics would instead be detectable at smaller Reynolds numbers, as suggested by the scaling relation $D/\lambda \sim D/(\eta \sqrt{Re_{\lambda})}$.%(see \cite{Homann:arxiv:2009} for results at $Re_{\lambda} = 32$).

\begin{figure}%[h!]
\centerline{\includegraphics[width=7.0cm]{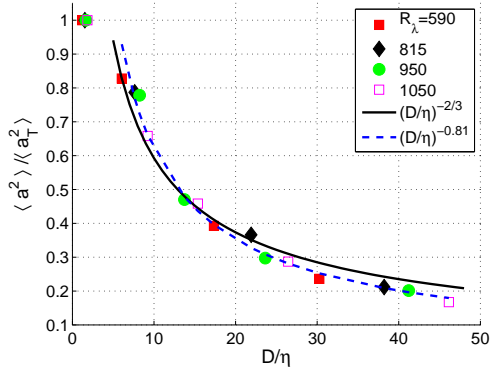}}
\caption{One-component normalized acceleration variance $\langle a_D^2 \rangle/\langle a_T^2 \rangle$ versus particle size. In order to be able to compare flows at varying Reynolds number $Re_{\lambda}$, the particle acceleration variance is normalized by the one measured with the smallest particles (tracers (T)) for which $D/\eta \leq 2$ at all $Re_{\lambda}$. Solid line: Kolmogorov scaling $\langle a_D^2 \rangle \propto (D/\eta)^{-2/3}$. Dashed line: refinement including intermittency corrections (section~3.4).}
\label{fig2}
\end{figure}

\subsection{Particle acceleration variance}\label{sec:var}
%%%%%%%%%%%%%%%%%%%%%%%%%%%%%
With one component of velocity probed by the eLDV system and in a situation where the direction of motion is not prescribed, the first moment of the distribution of acceleration is null. One expects that the second moment (acceleration variance)  reduces with increasing particle size, because the pressure forces which mainly cause the motion are averaged over a growing area. As shown in figure~\ref{fig2}, this is indeed observed. The evolution of the acceleration variance measured here is in agreement with previous studies by~\cite{Voth:2002,qureshi:2007,Voth:2009}: when normalized by the acceleration variance of the smallest particles (fluid tracers, noted $ \langle a^2_T \rangle$), the quantity  $\langle a_D^2 \rangle  / \langle a^2_T \rangle$ exhibits a decrease consistent with the power law $(D/\eta)^{-2/3}$ for all the Reynolds numbers and inertial range particle sizes. Let us recall that this power-law behavior is obtained when one assumes that the particle acceleration scales like pressure increments over a length equal to the particle's diameter. In the inertial range of scales, this argument yields the scaling $\langle a_D^2 \rangle \propto \langle (\delta_D P / D)^2 \rangle \sim D^{4/3-2} = D^{-2/3}$ -- where $(\delta_D P)^n \equiv S^n_p(D)$ is the pressure $n^{\rm th}$-order structure function. We show later in text that one may also include intermittency corrections to obtain the dashed line shown in figure~\ref{fig2} which yields an improved agreement with measured data. 

\subsection{Particle response time}\label{sec:resp}
\begin{figure}%[t!]
	\includegraphics[width=7.0cm]{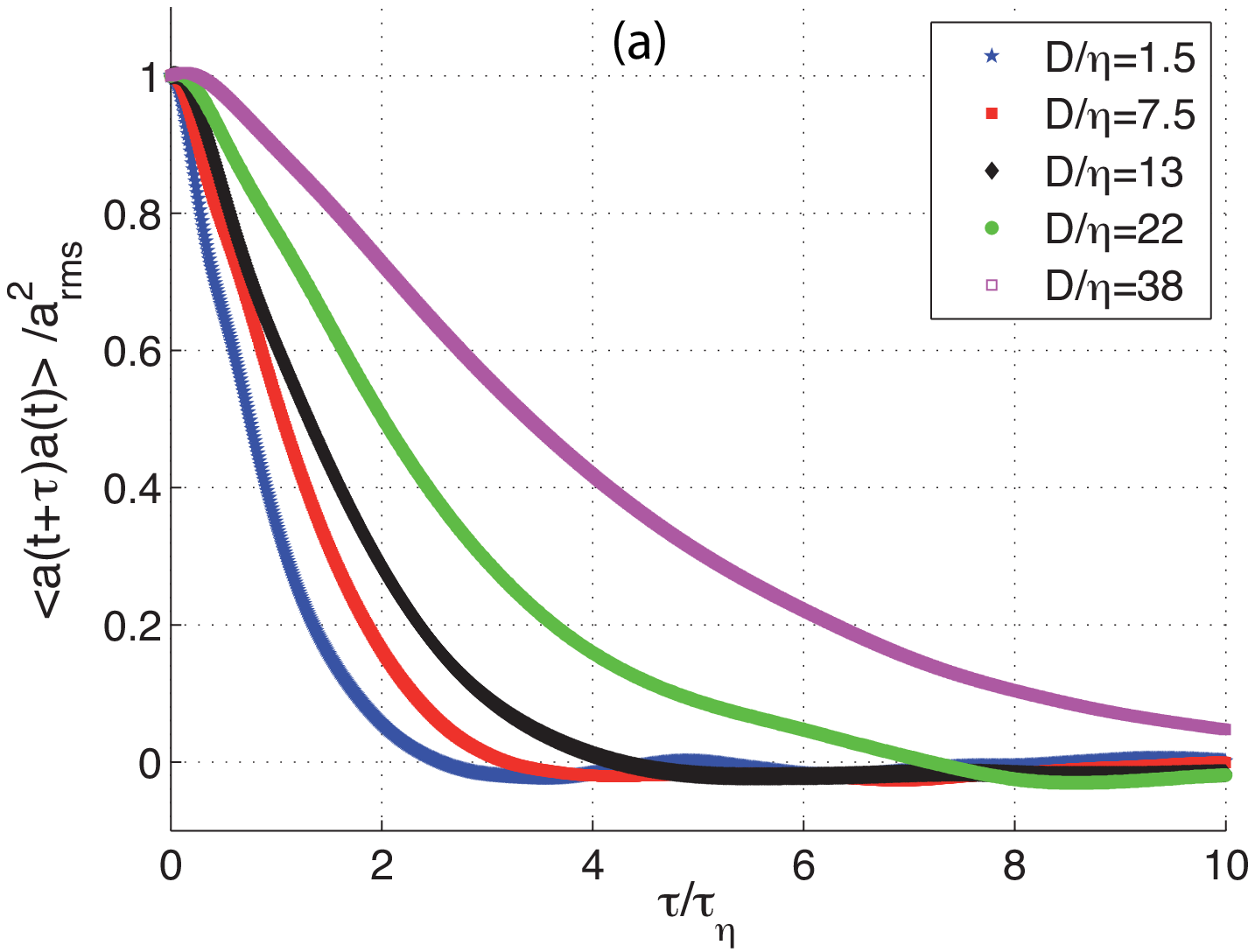}
   \includegraphics[width=7.0cm]{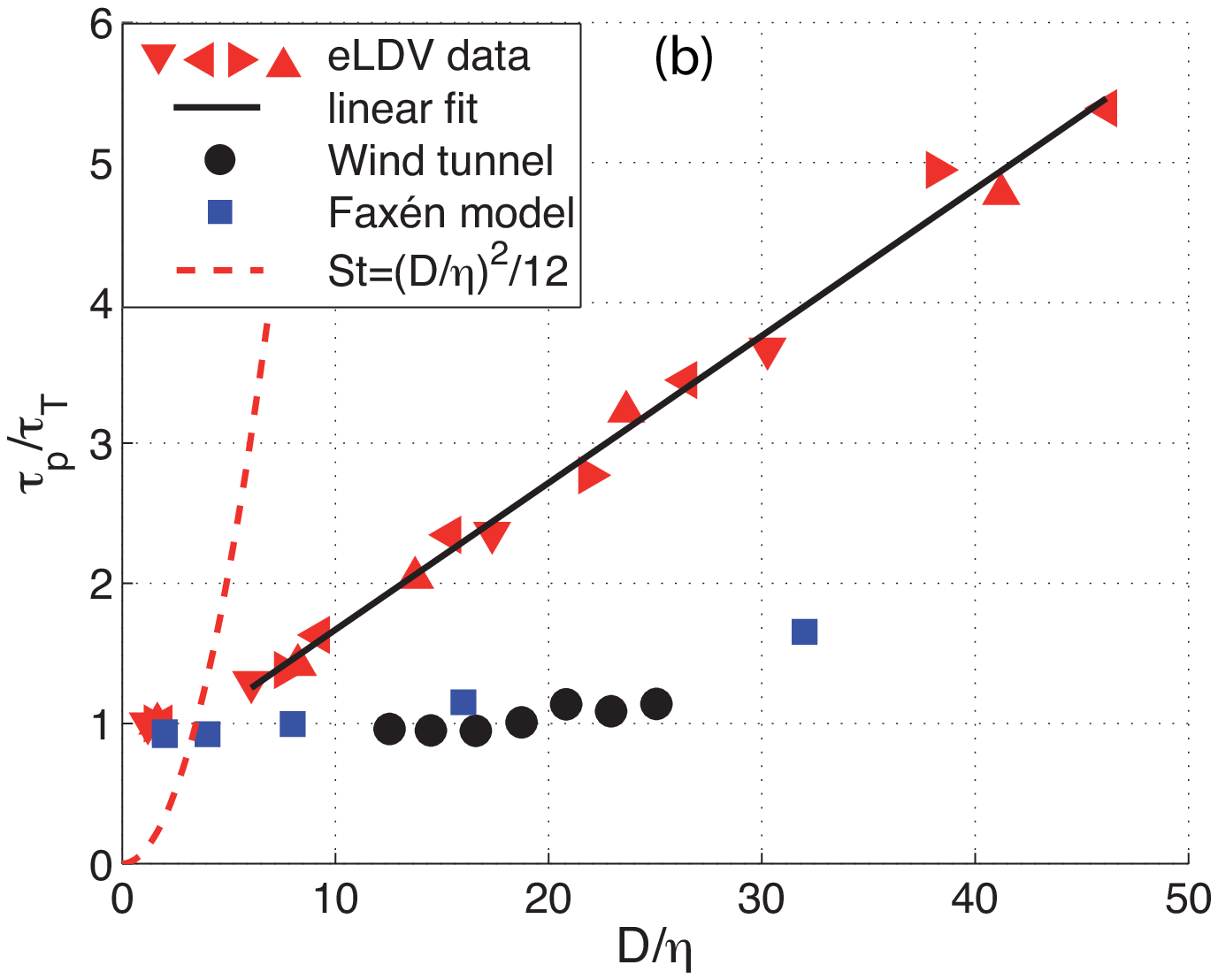}
   \caption{(a) Acceleration auto-correlation functions for $Re_{\lambda}=815$;  (b) evolution of particle response times. The wind tunnel data is extracted from~\cite{qureshi:2008}; the Fax\'en data are from~\cite{Calzavarini:JFM:2009}.  For the eLDV data the symbols corresponds to increasing Reynolds numbers: $(\triangledown) \; Re_{\lambda}=590$; $(\triangleleft) \; Re_{\lambda}=815$; $(\triangleright) \; Re_{\lambda}=850$; $(\Delta) \; Re_{\lambda}=1050$. }
   \label{fig7}
\end{figure}

%%%%%%%%%%%%%%%%%%%%%%%%%%%
A characteristic time for the evolution of a particle response to evolving flow conditions is obtained from the acceleration auto-correlation functions. Their shape and evolution with particle size is shown in figure~\ref{fig7}(a), {for $Re_{\lambda} = 815$}. In agreement with previous observations for tracers, the auto-correlation for small particles vanishes in times of the order of a few Kolmogorov time $\tau_\eta=\sqrt{\nu/\epsilon}$. As could be expected, the response time $\tau_p$, as defined as the integral over time of the positive part of $C_{aa}(\tau)$, increases with size, at any given Reynolds number. Our observation is that for a given Reynolds number, $\tau_p$ increases {\it linearly} with the particle diameter $D$ for sizes larger than about $10\eta$.  In addition, as shown in figure~3(b)(red/triangle symbols), measurements performed at various $Re_{\lambda}$ all line-up on the same curve, confirming that the evolution is indeed given by the relevant dimensionless variables, $\tau_p/\tau_\eta = f(D/\eta)$, {\it i.e.} when the response time is counted in units of the Kolmogorov time $\tau_\eta=\sqrt{\nu/\epsilon}$ and the particle size counted in units of dissipative scale $\eta=(\nu^3/\epsilon)^{1/4}$. 

%-- in the experiment, the dissipation $\epsilon$ is estimated from the power consumption of the motors, an estimation which is proportional to the local dissipation at the center of the VK flow, as shown in~\cite{Tabeling:PRE:1996,Mordant:JP2:1997}. 

The observed behavior is quite different from the prediction of point-particle (PP) models, for which the Stokes drag term becomes rapidly negligible when the particle size increases, so that the response time remains that of fluid tracers \cite{volk:2008b}.  A first refinement of the PP-model is to account for size effects by averaging the flow fields over the area of the particle (for the estimation of drag) and over its volume (for added mass effects); this is the essence of the Fax\'en-corrected model (FC) introduced by~\cite{Calzavarini:JFM:2009}. Using it, the authors have observed a variation of the particle response time with size:  it increases by almost a factor of 2 when the size of the particle increases from $D=2\eta$ to  up to $D=32\eta$. This finding was in general agreement with experimental measurement in a wind tunnel by~\cite{qureshi:2008}. As shown in figure~3(a), our observations in a von K\'arm\'an flow show a much steeper increase: the response time of the particles is about four times that of the tracers when the diameter has grown to $32\eta$, and the variation is roughly linear. However, our data are not sufficiently extended to exclude the scaling $\tau_p/\tau_\eta = (D/\eta)^{2/3}$, which comes from assuming that the response time varies as the eddy turnover time  at the scale of the particle.

\subsection{Particle acceleration probability density function}\label{sec:flat}
\begin{figure}
\centerline{\includegraphics[width=8.0cm]{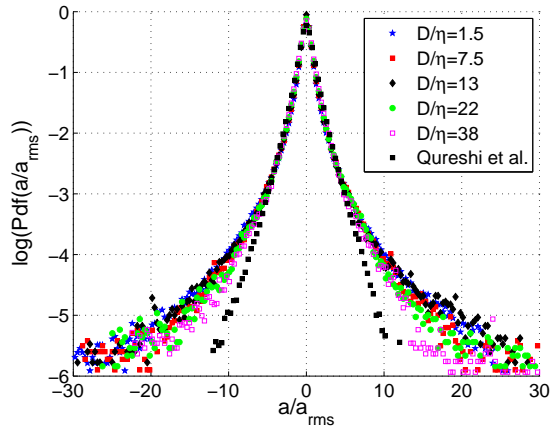}}
\caption{PDFs of particle acceleration at $Re_{\lambda} = 815$, normalized to the variance. The wind tunnel data by~\cite{qureshi:2007,qureshi:2008} is at $Re_{\lambda}=160$.}
\label{fig3}
\end{figure}

%%%%%%%%%%%%%%%%%%%%%%%%%%%%%%%%%%%%%%
The estimation of higher even moments of particle accelerations requires specific data processing, as we show in the following. We first discuss the raw distributions of the acceleration. In figure~\ref{fig3}, they are shown for $Re_{\lambda} = 815$; the particles accelerations have been normalized to their variance (whose behavior has been discussed above).  

To leading order, the distribution functions are very similar, as observed in the wind tunnel measurements by~\cite{qureshi:2007,qureshi:2008}. There is no reduction to Gaussian statistics as the particle size grows well into the inertial range (see also~\cite{Gasteuil:2009} with particles on integral size). However, as can be observed in the figure, there is a significant influence of the flow generation process on the form of the probability distribution. In figure~\ref{fig3}, the PDF for the smallest particles is identical to that observed for tracers in the VK flow operated by Bodenschatz's group, {\it c.f.}~\cite{Voth:2002}, and in numerical simulations~\cite{yeung:2002,mordant:2004}. It is different from the PDFs reported by~\cite{qureshi:2007,qureshi:2008} and by~\cite{ayyala:2006} from wind-tunnel measurements.  These observed variation are  more pronounced than what could be expected from Reynolds number variations alone between the experiments, and we thus propose that the PDFs of accelerations are not universal, but flow-dependent. This conclusion is linked to the observation (reported in~\cite{Ouellette:NJP:2006}) that the Lagrangian small scale dynamics still reflects the anisotropy of the larges scales. 

\begin{figure}
\centerline{\includegraphics[width=12.0cm]{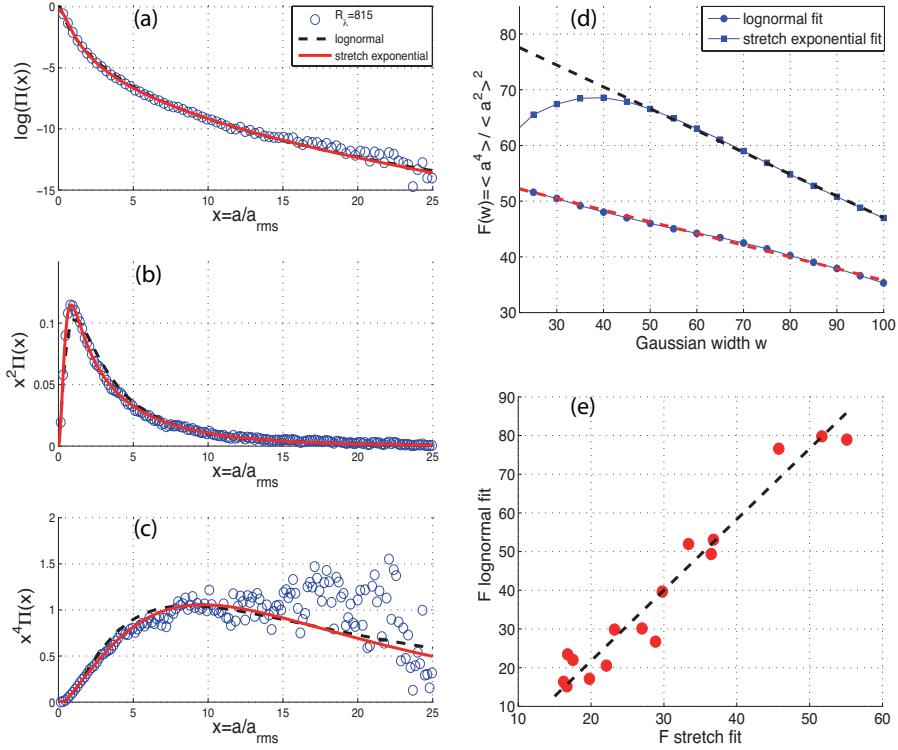}}
\caption{(left): lognormal and stretch exponential fits $\Pi(x=a/a_{\rm rms})$ and its moments $x^2\Pi(x)$ and $x^4\Pi(x)$ , for data at $Re_{\lambda} = 815$ processed using a gaussian smoothing with width $w=20$. ($\circ$) filtered experimental data, ($-$) stretch exponential fit (with flatness $F=55$), ($--$) lognormal fit with ($s=0.96$ and $F=68$). (right, top): evolution of the flatness estimated from the fitting functions, as a function of $w$, and corresponding linear interpolations to zero width. (right, bottom): relative evolution of the two estimators. }
\label{fig4}
\end{figure}

Investigating the possibility of sub-leading changes in the statistics of acceleration with size or Reynolds number can be done by studying higher order moments, starting with the distribution flatness. It requires a converged measurement of the PDFs and, as shown for tracer particles by~\cite{Mordant:PhyD:2004}, this implies extremely large data sets. As a first attempt, we fit the acceleration PDF with a model functional form which we then use to {\it estimate} the flatness of the distribution. The procedure is as follows: we assume that the statistics is described by a functional form ${\cal F}_{\theta}(a)$; $\{\theta\}$ is a set of adjustable parameters which are determined by minimizing the distance $a^2\Pi(a) - a^2 {\cal F}_{\theta}(a)$, where $\Pi(a)$ is the measured distribution. Two trial distributions have been tested: 

$$
{\cal F}^{\rm LN}_{s}(a) = \frac{e^{3s^2/2}}{4\sqrt{3}} \left( 1 - {\rm erf} \left( \frac{\ln |a/\sqrt{3}| + 2 s^2}{s\sqrt{2}} \right)  \right) 
$$

which stems from the assumption that the acceleration amplitude has a lognormal distribution, and the stretched exponential functional form

$${\cal F}^{SE}_{s}(a) = A \text{exp}\left( \frac{-a^2}{2\sigma^2\left(1+\left | \frac{a \beta}{\sigma} \right |^\gamma\right)}\right)$$

($A$ being a normalization constant) which has three adjustable parameters $(\alpha,\beta,\gamma)$, and allows for a finer adjustment of the distribution in the tails. Note that with distributions having such extended wings, a `brute force'  measurement of the flatness factor within a 5\% accuracy would mean a resolution of the distribution up to about 100 standard deviations, and events with probability below $10^{-11}$ --  clearly outside of direct experimental reach!

Figure \ref{fig4} shows a comparison of the acceleration pdf for the smallest particles at $Re_\lambda=815$ and the corresponding fits that minimize the distance to the quantity $a^2\Pi(a)$. As one can see, both functional form fit correctly the experimental data up to $a/a_{rms} \sim 20$, the stretch exponential form showing a better agreement with the second order moment. As reported in previous studies, the moments of the acceleration PDFs strongly depend on the width $w$ of the smoothing kernel used to extract the velocity data from the modulated optical signal (cf. section~2). We thus estimate the flatness by fitting the different PDFs for decreasing $w$ and then interpolate to zero width. The result of this procedure is shown in figure~\ref{fig4}(d) for the two trial functional forms. As can be seen the flatness derived from the lognormal estimator is roughly $1.5$ higher than the one computed from the stretch exponential estimator. Indeed, we find that the flatness computed from each function follow an identical behavior:  when plotting, as in figure~\ref{fig4}(e), the values estimated from the lognormal fit versus the ones computed from the stretch exponential estimator, one observe a linear dependance. The values of the flatness reported are understood as estimates, the true values strongly depending on the real shape of the acceleration PDF in the far tails.  We propose however that the {\it variations} detected here yield a first order approximation of the evolution of acceleration statistics with particle size.  

\begin{figure}
\centerline{\includegraphics[width=6.0cm]{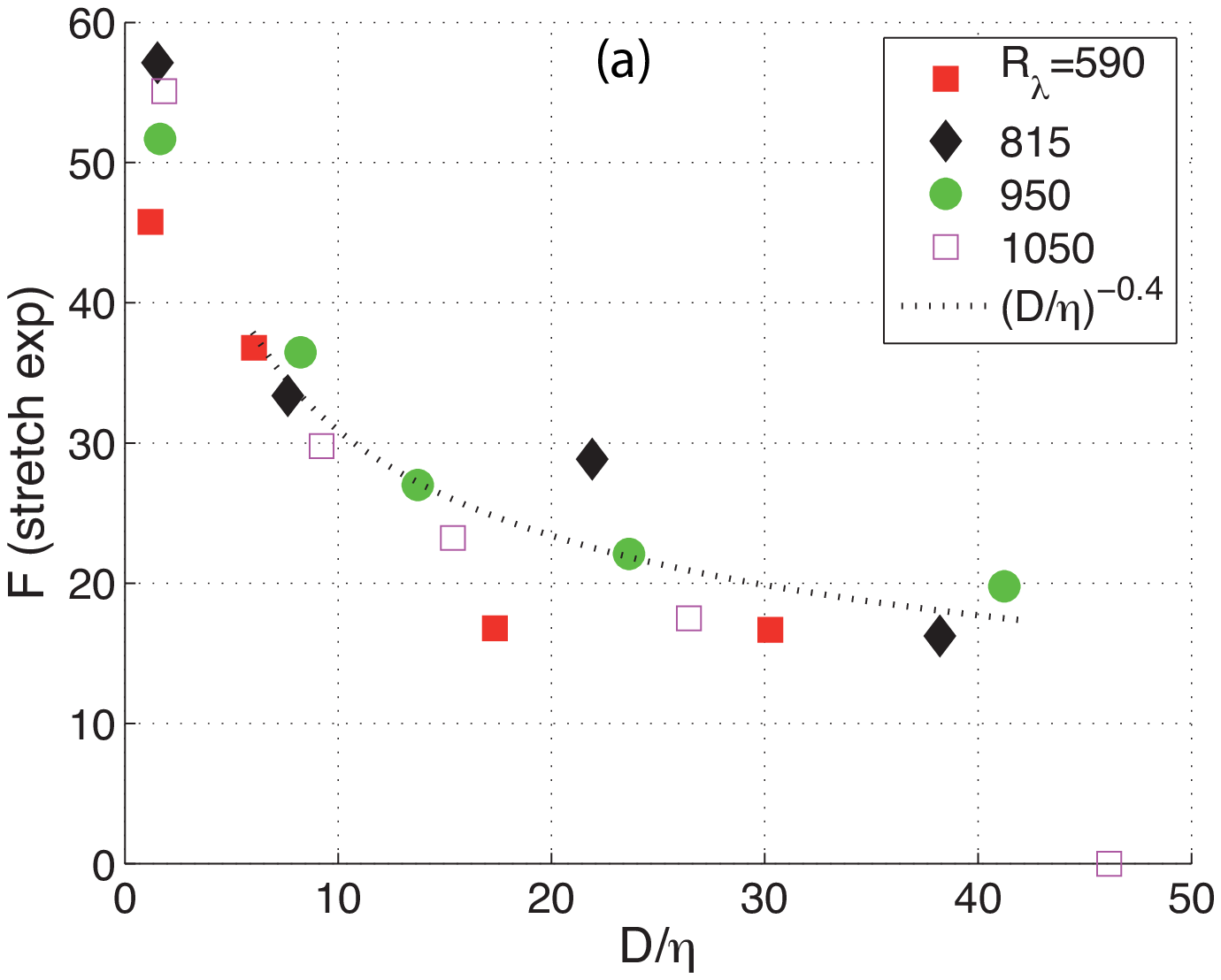}
\includegraphics[width=6.0cm]{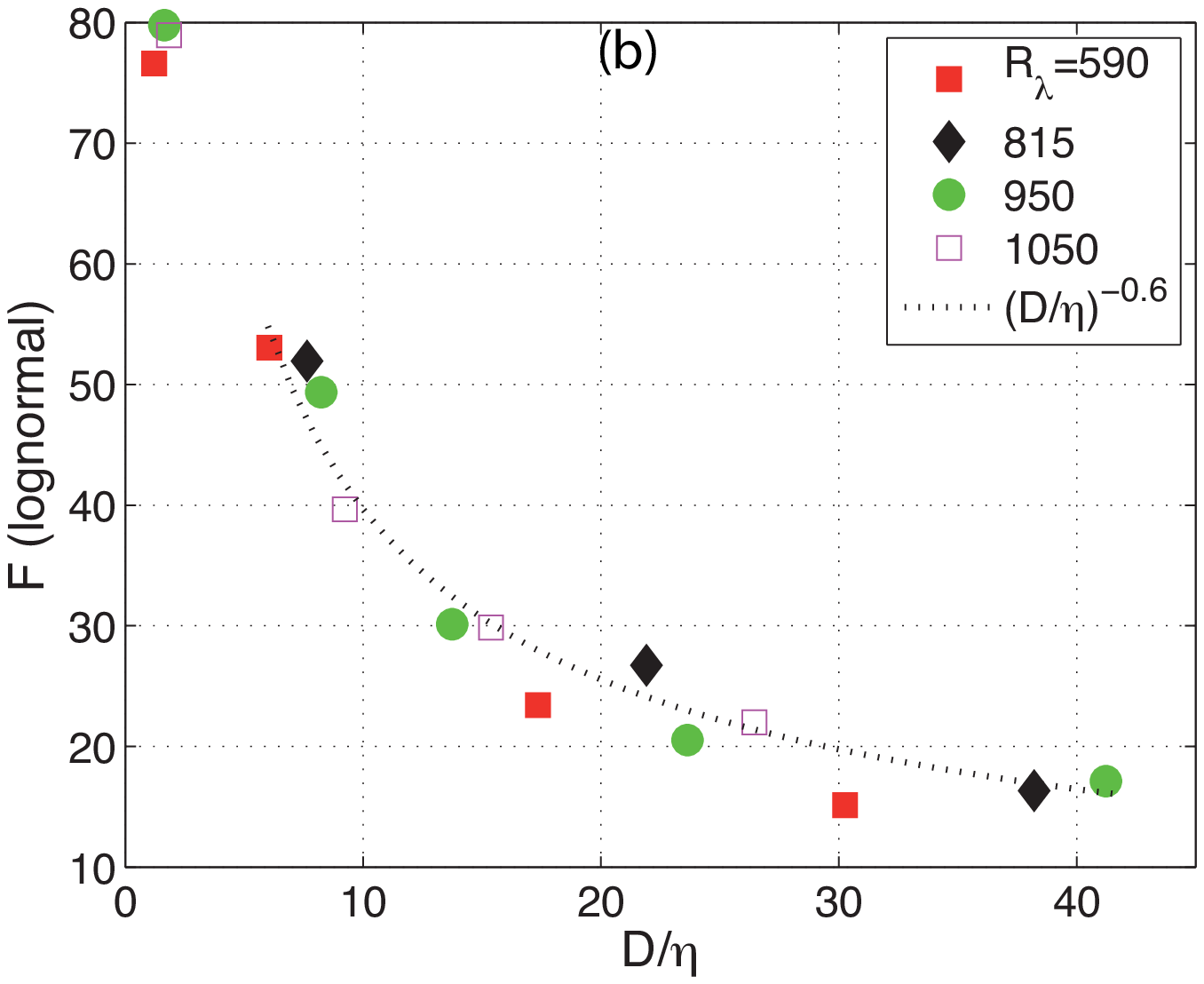}}
\caption{Variation of the estimated flatness as a function of $D/\eta$ for different sizes and different reynolds numbers.  (a) stretch exponential estimator; (b) lognormal estimator.}
\label{fig6}
\end{figure}

The results are shown in Figure \ref{fig6}(a,b). For both estimators, one observes a reduction of the flatness with increasing particle size. The observed decrease can be understood if one takes into account the intermittency of the pressure increments for inertial range separations. Indeed, following the approach developed by~\cite{Voth:2002,qureshi:2007}, one estimates the acceleration flatness $F(D)$ by assuming that the force acting on the particles is dominated by the pressure gradient at size $D$. All moments of the acceleration ($\langle a_D^n \rangle$) should then scale as $S_P^n(D)/D^n$, its behavior being dictated by the pressure structure functions.
Now, in order to estimate the pressure structure functions, one can either use the ansatz that pressure increments scale as the square of velocity increments, $\langle \delta_D P \rangle \propto \langle (\delta_D v)^2 \rangle$ or one can measure directly the scaling of pressure in the experiment. In the first case, one obtains $S_P^n(D) \propto D^{2\zeta_n}$, $\zeta_n$ being the structure function exponents of the velocity increments. One then has $F(D) \propto  D^{\zeta_8 - 2 \zeta_4} \sim D^{-0.42}$, if one assumes a lognormal scaling for the Eulerian velocity structure functions as in~\cite{Chevillard:PHYD:2006}, independent of the Reynolds number.
The direct experimental measurement of pressure (using piezoelectric sensor mounted flush with the lateral wall, in the mid-plane of the flow) yields {$S_P^2 \propto D^{1.2 \pm 0.1}$ and $S_P^4/(S_P^2)^2 \propto D^{-0.38 \pm 0.03}$, values which are in agreement with Eulerian DNS data at $Re_{\lambda} =180$ (from the DNS by  ~\cite{Calzavarini:JFM:2009}).
In the case of the stretch exponential estimator, these predictions for the scaling exponent are consistent with the the value $\alpha \sim -0.4$ obtained by fitting the data with a power law $F_s(D)=A(D/\eta)^\alpha$ with $D/\eta$ in the range $\left[10, 40\right]$. In the case of the lognormal estimator one finds a scaling law $F_l(D) \propto (D/\eta)^{-0.6}$. Here we stress that ${\cal F}^{\rm LN}_{s}(a)$ and ${\cal F}^{SE}_{s}(a)$ are intrinsically different distributions, therefore it is in principle impossible to fit both curves with the same scaling exponent. The true value of the exponent (if it exists) should depend on the real shape of the PDFs. However, the consistency between the estimation using the stretch exponential estimator and the Eulerian measurements of pressure suggests that this estimator is a good fit of the acceleration PDFs in the case of large acceleration flatness ($F>20$). This was confirmed by comparing the quality of the two different estimators with data obtained at $Re_{\lambda} =180$ ($F \simeq 27.5$, \cite{Calzavarini:JFM:2009}). Using a troncated dataset as a test, the stretch exponential estimator showed to be a better fit than the lognormal estimator, and was able to give an estimate of the Flatness $15\%$ lower than the converged value computed with the all dataset.}
{Finally, we note that following the same approach, one can use the second order pressure structure function to get a new estimate of the decrease of the acceleration variance $\langle a_D^2 \rangle$. Assuming a lognormal scaling for the velocity increments, one then finds $\langle a_D^2 \rangle \propto \langle (\delta_D P / D)^2 \rangle \sim D^{\zeta_4-2} = D^{-0.78\pm0.02}$ very close to the experimental measurements which yields  $\langle a_D^2 \rangle \sim D^{-0.8\pm0.1}$. These two values are in a very good agreement with the best fit shown as a dashed line in figure~2 which yields $\langle a_D^2 \rangle/\langle a_T^2 \rangle \propto (D/\eta)^{-0.81}$. In our opinion this is an indirect proof that intermittency should play a role, and that our results concerning the acceleration flatness, if preliminary, are consistent.}

\section{Concluding remarks}\label{sec:end}
%%%%%%%%%%%%%%%%%%%%%%%
Scaling properties of the acceleration of (neutrally buoyant) inertial particles, {\it i.e.} concerning finite size effects for the motion of advected particles, have been investigated in detail. One finding is statistics is well accounted for by the behavior of pressure increments, which in turn are connected to the Eulerian properties of velocity increments. A first estimation concerning the evolution of the particle acceleration PDF with size suggests a reduction of the flatness with a $(D/\eta)^{-0.4}$ scaling behavior. However, since the precise functional form actually depends on the large scale flow properties, this results need to be confronted to measurements in other flows, such as grid or jet turbulence. Another observation is that the evolution of the acceleration auto-correlation functions allow for the estimation of a particle response time. Our finding is that it increases linearly with the particle size in the range of sizes explored. One point that will deserve further studies is the dependence on Reynolds number. Here we have achieved a satisfactory collapse of measured quantities when the particle size is made non-dimensional using the Kolmogorov length $\eta$. This may be justified for the $R_\lambda$ values and particle sizes investigates here; for larger particles or lower Reynolds numbers one may also want to probe scaling properties as a function of $D/L$, with $L$ the flow integral scale. Here, $L$ was much larger than the largest particles size, but such may not be the case for all studies. \\

{\bf Acknowledgements}  The authors thank Mickael Bourgoin for many useful discussions. Work supported by $ANR-BLAN-07-1-192604$.  \\

%\bibliographystyle{jfm2}
%\bibliography{biblio_fsm_2}

\end{document}